\documentclass[reqno]{amsart}
\usepackage{amsmath,amssymb,euscript,graphicx}
\usepackage[arrow,matrix,curve]{xy}
\numberwithin{equation}{section}

\newcommand\bZ{\mathbb{Z}}
\newcommand\bR{\mathbb{R}}
\newcommand\bC{\mathbb{C}}
\newcommand\be{\mathfrak{e}}
\newcommand\bm{\mathfrak{m}}

\newcommand\Hom{\operatorname{Hom}}
\newcommand\Ext{\operatorname{Ext}}
\newcommand\coker{\operatorname{coker}}
\newcommand\cc{\mathrm{cc}}
\newcommand\im{\operatorname{im}}
\renewcommand\Re{\operatorname{Re}}
\renewcommand\Im{\operatorname{Im}}

\newcommand\gam\gamma
\newcommand\del\delta
\newcommand\sig\sigma
\newcommand\om\omega
\newcommand\Sig\varSigma

\newcommand\cB{\EuScript{B}}
\newcommand\cH{\EuScript{H}}
\newcommand\cM{\EuScript{M}}
\newcommand\cN{\EuScript{N}}
\newcommand\MH{\cM_{\mathrm{H}}}
\newcommand\Mf{\cM_\mathrm{flat}}

\newcommand\tSig{{\tilde\Sig}}
\newcommand\bSig{{\partial\Sig}}
\newcommand\LG{{}^L\!G}
\newcommand\ZG{{Z(G)}}
\newcommand\pG{\pi_1(G)}
\newcommand\tG{{\widetilde G}}
\newcommand\sfrac[2]{\raisebox{-.1ex}{\scalebox{1.1}[1.2]{$\frac{#1}{#2}$}}}
\newcommand\map[1]{\stackrel{{#1}}\longrightarrow}

\begin{document}

\hfill {\tt arXiv:1811.nnnnn\,[hep-th]}\\
${}\,$ \hfill to appear in the Proceeding of ICHEP2018

\bigskip

\title{Testing $S$-duality with non-orientable surfaces}
\maketitle

\vspace{-1em}

\begin{center}
{\large Siye Wu}\\
\bigskip
{\small Department of Mathematics, National Tsing Hua University,
Hsinchu 30013, Taiwan\\
E-mail address: {\tt swu@math.nthu.edu.tw}}
\end{center}

\begin{abstract}
Kapustin and Witten showed that a twisted version of $N=4$ gauge theory
in four dimensions compactifies to a two-dimensional sigma-model whose
target space is the Hitchin moduli space.
In this talk, I consider the reduction of the gauge theory on a four
dimensional orientable spacetime manifold which is not a global product of
two surfaces but contains embedded non-orientable surfaces.
The low energy theory is a sigma-model on a two dimensional worldsheet whose
boundary components end on branes constructed from the Hitchin moduli space
associated to a non-orientable surface.
I will also compare the discrete topological fluxes in four and two
dimensional theories and verify the mirror symmetry on branes as predicted
by the $S$-duality in gauge theory.
This provides another non-trivial test of $S$-duality using reduction along
possibly non-orientable surfaces.
Finally, I consider the quantisation of the Hitchin moduli space from a
non-orientable surface as an example of quantisation via branes and mirror
symmetry.
\end{abstract}

\section{Topological sectors of $4$-dimensional gauge theory}
We consider $4$-dimensional gauge theory with a gauge group $G$ which is a
connected compact semisimple Lie group.
Electric charges of a fundamental particle (or field) are irreducible
representations of $G$, classified by the weight vectors in the positive
Weyl chambre.
On the other hand, magnetic monopoles, in classical gauge theory, are from
gauge potentials (or connections) that are singular in space.
Given a homomorphism from $\mathrm U(1)$ to $G$, there is a Yang-Mills
connection induced by the Dirac monopole in $\mathrm U(1)$ gauge theory
of charge $1$.
Therefore the magnetic charges are elements in $\Hom(\mathrm U(1),G)$ up to
conjugations in $G$ (to account for gauge equivalence).
The latter corresponds to the coweight vectors in the positive chambre.
At the quantum level, magnetic charges describe collective excitations
whereas electric charges are of elementary excitations.

In addition to these charges, a $4$-dimensional gauge theory has discrete
fluxes which belong to finite Abelian groups.
After Wick rotation, the spacetime has Euclidean signature and can be chosen
as a compact orientable $4$-manifold $X$.
Classical gauge theory builds upon the geometry of principal $G$-bundles over
$X$.
These bundles are classified topologically by characteristic classes in
$H^4(X,\pi_3(G))$ and $H^2(X,\pG)$.
When $G$ is simple, such as $\mathrm{SU}(n)$ with $n\ge2$, we have
$\pi_3(G)\cong\bZ$, and $H^4(X,\pi_3(G))\cong\bZ$ contains the instanton
numbers.
Since $\pG$ is finite Abelian, the group $H^2(X,\pG)$ is necessarily torsion
and it contains the discrete fluxes of 't~Hooft \cite{tH79}.

When $X$ has a splitting of space and time, i.e., when $X=T^1\times Y$, where
$T^1$ is a circle in the time direction and $Y$ is a compact orientable
spatial $3$-manifold, there is a decomposition
\[ H^2(X,\pG)\cong H^2(Y,\pG)\oplus H^1(Y,\pG). \]
An element $m\in H^2(Y,\pG)$ classifies the topology of the $G$-bundle over
a time slice $Y$ and is called the discrete magnetic flux.
A fixed $m$ determines a well defined sector both classically and quantum
mechanically.
On the contrary, an element $a\in H^1(Y,\pG)\cong H^1(Y,H^1(T^1,\pG))$
contains the information of the entire time interval and can not be fixed
consistently.

Instead, we interpret discrete electric fluxes as the momenta of discrete
translations on field configurations when the centre $\ZG$ of $G$ is
non-trivial.
Recall that $\ZG$ is also a finite Abelian group.
We claim that $H^1(Y,\ZG)$ is a discrete symmetry in the gauge theory.
An element $g\in H^1(Y,\ZG)=\Hom(\pi_1(Y),\ZG)$ modifies the holonomy of
a connection $A$ along a loop $\gam$ in $Y$ by $g([\gam])\in\ZG$, where
$[\gam]\in\pi_1(Y)$ is the class represented by $\gam$ in the fundamental
group.
This procedure preserves the curvature (or field strength) and hence the
classical action.
The quantum Hilbert space is then a representation of $H^1(Y,\ZG)$ and
decomposes according to the types of its irreducible representations.
So the quantum theory consists of sectors labelled by $e\in H^1(Y,\ZG)^\vee$,
where for any Abelian group $A$, the Pontryagin dual  
$A^\vee:=\Hom(A,\mathrm U(1))$ is the group of characters of $A$.
The elements $e$ are the discrete electric fluxes.

When $G$ is exchanged with its Langlands dual $\LG$ or its magnetic group,
so do the weight and coweight lattices and the sets of electric and magnetic
charges \cite{GNO}. 
In adition, we have
\[ H^1(Y,\ZG)^\vee\cong H^2(Y,\pG),\qquad H^2(Y,\pG)\cong H^1(Y,Z(\LG))^\vee \]
from Poincar\'e duality.
So the discrete electric fluxes in the $G$-theory are the discrete magnetic
fluxes in the $\LG$-theory, and vice versa.
This is consistent with the electric-magnetic duality (or $S$-duality)
proposed by \cite{MO}.

\section{Reduction to $2$-dimensions along orientable surfaces}\label{sec:C}
In \cite{KW}, Kapustin and Witten considered the reduction of a twisted $N=4$
supersymmetric gauge theory on the $4$-manifold $X=\Sig\times C$ to a
sigma-model whose worldsheet is $\Sig$ along a compact orientable surface $C$
of genus $g(C)>1$.
The target space of the low energy theory is the Hitchin moduli space
$\MH(C,G)$, which is a hyper-K\"ahler manifold with complex structures $I,J,K$
and K\"ahler forms $\om_I,\om_J,\om_K$ following the notations of \cite{KW}.

When the worldsheet has a splitting $\Sig=T^1\times S^1$, so does the
$4$-dimensional spacetime $X=T^1\times Y$, in which $Y=S^1\times C$.
In gauge theory, we write the discrete magnetic flux $m=m_0+m_1$ and the
discrete electric flux $e=e_1+e_0$ according to the decompositions
\[ H^2(Y,\pG)\cong H^2(C,\pG)\oplus H^1(C,\pG),\quad
   H^1(Y,\ZG)^\vee\cong H^1(C,\ZG)^\vee\oplus H^0(C,\ZG)^\vee.  \]
In the sigma-model, $m_0\in\pi_0(\MH(C,G))$ labels the connected component in
which the string propagates while $m_1\in\pi_1(\MH(C,G))$ is the winding of
the string.
On the other hand, $e_1$ is a discrete momentum of the symmetry group
$H^1(C,\ZG)$ acting on $\MH(C,G)$, whereas $e_0$ labels the flat $B$-field
on $\MH(C,G)$ coupled to the sigma-model \cite{KW}.

Electric-magnetic duality in four dimensions, which is believed to be exact
in $N=4$ gauge theories, reduced to mirror symmetry in two dimensions
\cite{BJSV,HMS,KW}.
Indeed, $\MH(C,G)$ and $\MH(C,\LG)$ are mirrors to each other \cite{HT,DP12}
in the sense of \cite{SYZ}.
With the exchange of $G$ and $\LG$, the roles of $m_0$ and $e_0$,
$m_1$ and $e_1$, also interchange.
The mirror correspondence of branes on $\MH(C,G)$ and $\MH(C,\LG)$
explains much of the geometric Langlands programme \cite{KW}.

\section{Reduction along possibly non-orientable surfaces}\label{sec:C'}
In \cite{Wu18}, we consider the gauge theory on a $4$-manifold which is
not a global product but contains embedded non-orientable surface $C'$.
Let $\pi\colon C\to C'$ be the orientation double cover:
there is a free $\bZ_2$ action on $C$ and the quotient is $C'$.
For example, $C=S^2$ and $C'=C/\bZ_2=\bR P^2$.
But we will assume $C'$ is a connected sum of $g(C')>2$ copies of $\bR P^2$.
For the worldsheet, we take an orientable surface $\tSig$ with an
orientation-reversing $\bZ_2$-action.
The quotient $\Sig=\tSig/\bZ_2$ is a surface whose boundary $\bSig$ is the
fixed-point set of $\bZ_2$ on $\tSig$, which is assumed to be non-empty.
For example, $\Sig$ is a disc for $\tSig=S^2$ with a reflection along the
equator.
The $4$-manifold $X=\tSig\times_{\bZ_2}C$ (quotient by the diagonal action)
is a smooth orientable $4$-manifold without boundary \cite{Wu18}.
There is a projection $\pi_X\colon X\to\Sig$ by forgetting $C$.
The inverse image $\pi_X^{-1}(\sig)$ is a copy of $C$ if $\sig$ is in the
interior of $\Sig$ but is $C'$ if $\sig\in\bSig$.
So $X$ contains a $\bSig$-family of non-orientable surfaces $C'$.

Hitchin's equations make sense on a non-orientable surface $C'$ and the
moduli space $\MH(C',G)$ of solutions modulo gauge equivalence is introduced
and studied in details in \cite{HWW}.
There is a map $p\colon\MH(C',G)\to\MH(C,G)$ by pulling back fields from $C'$
to $C$.
The image $\cN(C,G)$ of $p$ is contained in the $\bZ_2$-invariant part
$\MH(C,G)^{\bZ_2}$ of $\MH(C,G)$; the latter is Lagrangian in $\om_I,\om_K$
and holomorphic in $J$.
On the smooth part, $p\colon\MH(C',G)\to\cN(C,G)$ is a finite regular
$\ZG_{[2]}$-cover.
Here $A_{[2]}=\{a\in A:2a=0\}$ is the $2$-torsion subgroup for any Abelian
group $A$.

In the limit of large $\tSig$ (or $\Sig$) and small $C$ (or $C'$), the same
$N=4$ gauge theory on $X$ reduces to a sigma-model of target space $\MH(C,G)$
on $\Sig$ with the boundary $\bSig$ living on branes \cite{Wu18}.
Among the bosonic variables are the maps $u\colon\Sig\to\MH(C,G)$ and
$u'\colon\bSig\to\MH(C',G)$ satisfying $p\circ u'=u|_\bSig$.
For each $e_2\in(\ZG_{[2]})^\vee$, we get a flat Chan-Paton bundle
$\ell^{e_2}=\MH(C',G)\times_{e_2}\bC$ over $\cN(C,G)$.
On the other hand, there is a decomposition
$\MH(C',G)=\bigsqcup_{m_2\in\pG/2\pG}\MH^{m_2}(C',G)$ according to
the topological types of $G$-bundles over $C'$, and we let
$\cN^{m_2}(C,G)=p(\MH^{m_2}(C',G))$.
Then $\cB^{e_2,m_2}=(\cN^{m_2}(C,G),\ell^{e_2})$ is a brane on $\MH(C,G)$
of type $(A,B,A)$.
The low energy theory on $\Sig$ contains sectors whose boundary conditions
are defined by $\cB^{e_2,m_2}$.

Consider the case $\tSig=T^1\times S^1$ with $\bZ_2$ acting on $S^1$ by
reflection.
Then $\Sig$ is a cylinder with two time-like boundary circles.
The sigma-model is about the propagation of an open string whose boundary
points are constrained on branes.
By homotopy calculations \cite{Wu18}, the relative winding of the open string
is $m_1\in H^1(C,\pG)/\pi^*H^1(C',\pG)$.
The presence of branes reduces the $H^1(C,\ZG)$ symmetry to $\pi^*H^1(C',\ZG)$
and the discrete momenta are $e_1\in\pi^*H^1(C',\ZG)^\vee$.
So the sectors of the $2$-dimensional theory are labelled by $m_1,e_1,m_2,e_2$.
The absence of $m_0$ is because only the component $\MH^{m_0=0}(C,G)$ supports
the branes whereas the absence of $e_0$ is due to an anomaly-free condition
\cite{Wu18} much like the Freed-Witten condition \cite{FW} for untwisted
strings.

The $4$-dimensional spacetime is $X=T^1\times Y$, where $Y=S^1\times_{\bZ_2}C$
is a smooth compact orientable $3$-manifold without boundary.
The sets of discrete electric and magnetic fluxes in the gauge theory are
given by the exact sequences \cite{Wu18}
\[ 0\to\sfrac{H^1(C,\pG)}{\pi^*H^1(C',\pG)}\to H^1(Y,\pG)
   \to(\pG/2\pG)^{\oplus2}\to0,   \]
\[ 0\to(\pi^*H^1(C',\ZG))^\vee\to H^1(Y,\ZG)^\vee
   \to(\ZG_{[2]}^{\oplus2})^\vee\to0.   \]
This matches the low energy data: we see the absence of $m_0,e_0$, the
relative windings $m_1$, the discrete momenta $e_1$, and two copies of
$m_2,e_2$ because an open string has two end points.

As usual, $S$-duality exchanging $G$ and $\LG$ reduces to mirror symmetry,
interchanging the roles of $m_1$ and $e_1$, $m_2$ and $e_2$.
The latter is made possible by the isomorphisms
\[ \sfrac{H^1(C,\pG)}{\pi^*H^1(C',\pG)}\cong(\pi^*H^1(C',Z(\LG)))^\vee,\qquad
   \pG/2\pG\cong(Z(\LG)_{[2]})^\vee,  \]
etc.
The brane $\cB^{e_2,m_2}$ in the original theory and ${}^L\!\cB^{m_2,e_2}$ in
the dual theory are related by a fibrewise Fourier-Mukai transform \cite{Wu18}
between branes on $\MH(C,G)$ and $\MH(C,\LG)$ that are dual special Lagrangian
fibrations.
This provides another non-trivial test of $S$-duality with dimensional
reduction along non-orientable surfaces, using the properties of the
moduli space $\MH(C',G)$.

\section{Adjustments for full generality}
The above consideration, while sufficient in many circumstances, is not yet
completely accurate in the most general setting.
It can happen that the electric and magnetic fluxes can not be simultaneously
fixed because the discrete symmetry $H^1(Y,\ZG)$ may change the topology of
the $G$-bundle over $Y$ \cite{Wu15,Wu18}.
More precisely, an element $g\in H^1(Y,\ZG)$ changes the discrete magnetic
flux $m$ to $m+\del(g)$, where $\del_Y$ is the connecting homomorphism in
the long exact sequence
\[ \cdots\to H^1(Y,Z(\tG))\to H^1(Y,\ZG)
\map{\del_Y}H^2(Y,\pG)\to H^2(Y,Z(\tG))\to\cdots, \]
where $\tG$ is the universal cover group of $G$.
This detail has been overlooked in the past literature because the map
$\del_Y$ is zero in many cases, such as when $H_1(Y)$ is torsion-free
(e.g., if $Y=S^1\times C$ as in \S\ref{sec:C}) or when the short exact
sequence $0\to\pG\to Z(\tG)\to\ZG\to0$ of $Z(\tG)$ splits.
But for the $3$-manifold $Y=S^1\times_{\bZ_2}C$ in \S\ref{sec:C'}, the map
$\del_Y$ can be non-zero.

A classical symmetry, if not anomalous at the quantum level, becomes part of
the automorphism group of the quantum operator algebra.
If the symmetry group does not act on the quantum Hilbert space, which is an
irreducible representation of the operator algebra, the symmetry is said to
be broken to the subgroup which does act on the Hilbert space.
Inner automorphisms are always in the unbroken subgroup, but typically, an
outer automorphism sends an irreducible representation to a different one.
In our case, since the Hilbert space is labelled by $m$, the unbroken subgroup
in $H^1(Y,\ZG)$ is $\ker(\del_Y)$, and consequently, the discrete electric
fluxes are in its character group $\be(Y,G):=\ker(\del_Y)^\vee$.
On the other hand, the discrete magnetic fluxes should label non-isomorphic
quantum theories, and they are in
$\bm(Y,G):=H^2(Y,\pG)/\im(\del_Y)=\coker(\del_Y)$.
For $Y=S^1\times_{\bZ_2}C$, the sets $\be(Y,G)$, $\bm(Y,G)$ of these rectified
fluxes can be computed explicitly and they are different from
$H^1(Y,\ZG)^\vee$, $H^2(Y,\pG)$.
Under $S$-duality, there are isomorphisms $\be(Y,G)\cong\bm(Y,\LG)$,
$\bm(Y,G)\cong\be(Y,\LG)$, exchanging the rectified electric and magnetic
fluxes \cite{Wu18}.

In the low energy theory, although each $\MH^{m_2}(C',G)$, $m_2\in\pG/2\pG$,
is expected to be connected, it is not preserved by the full covering group
$\ZG_{[2]}$ \cite{Wu15,Wu18}.
In fact, $g\in\ZG_{[2]}$ sends $m_2$ to $m_2+\del_{\bZ_2}(g)$, where
$\del_{\bZ_2}\colon\ZG_{[2]}\to\pG/2\pG$ is the connecting homomorphism of
a similar long exact sequence (replacing $Y$ by $B\bZ_2=\bR P^\infty$).
The map $\del_{\bZ_2}$ is zero when all elements in $Z(\tG)$ are of odd order
or when the above short exact sequence of $Z(\tG)$ splits.
So the branes are $\cB^{\bar e_2,\bar m_2}$, where $\bar e_2$ in
$\be(\bZ_2,G):=\ker(\del_{\bZ_2})^\vee$ defines a flat Chan-Paton line bundle
over the worldvolume $\cN^{\bar m_2}(C,G)$ that depends only on the cosets
$\bar m_2$ in $\bm(\bZ_2,G):=\coker(\del_{\bZ_2})$.
With the above modifications, the $4$- and $2$-dimensional data still match
just as in \S\ref{sec:C'}.
Under $S$-duality, there are isomorphisms $\be(\bZ_2,G)\cong\bm(\bZ_2,\LG)$,
$\bm(\bZ_2,G)\cong\be(\bZ_2,\LG)$.
So the twisting by Chan-Paton bundles on one side is mirror to displacements
of worldvolumes on the dual side.
With the rectified discrete fluxes, the mirror of $\cB^{\bar e_2,\bar m_2}$
is ${}^L\!\cB^{\bar m_2,\bar e_2}$.
We refer the reader to \cite{Wu18} for details.

\section{An example of quantisation via branes and mirror symmetry}
To quantise a symplectic manifold $(M,\om)$ via branes \cite{GW09}, one needs
a complexification $M^\bC$ with an anti-holomorphic involution fixing $M$.
There is also a holomorphic symplectic form $\om^\bC$ on $M^\bC$ such that
$\Re(\om^\bC)=\om$ on $M$.
The $\bZ_2$-action on $M^\bC$ lifts to a line bundle $\ell$ and preserves
its connection with curvature $\Re(\om^\bC)/\sqrt{-1}$.
We then have a space-filling coisotropic brane $\cB_\cc$ in the $A$-model on
$M^\bC$ with symplectic form $\Im(\om^\bC)$.
A trivial or flat line bundle on $M$ defines a Lagrangian $A$-brane $\cB_0$.
The quantisation of $(M,\om)$ is then $\Hom(\cB_\cc,\cB_0)$ \cite{GW09}.
In the $B$-model on the mirror of $M^\bC$, the quantum Hilbert space is
$\Ext(\cB_\cc^\vee,\cB_0^\vee)$, where $\cB^\vee$ is the dual of $\cB$
\cite{G}.

To quantise the Hitchin moduli space $\MH(C',G)$ with a non-orientable
surface $C'$, we need to generalise the above setting \cite{Wu18}:
$\MH(C',G)$ maps to $\MH(C,G)^{\bZ_2}$ in the complexification $\MH(C,G)$ by
a local diffeomorphism which is K\"ahler with respect to $\om_J$ \cite{HWW}.
A line bundle $\ell$ over $\MH(C,G)$ whose curvature is $\om_J/\sqrt{-1}$
defines $\cB_\cc$ \cite{Wu15}.
The quantisation of $\MH^{m_2}(C',G)$ is a sum of
$\cH^{\bar e_2,\bar m_2}=\Hom(\cB_\cc,\cB^{\bar e_2,\bar m_2})$ over
$\bar e_2\in\be(\bZ_2,G)$ in the $A$-model on $\MH(C,G)$ with $\om_K$.
By mirror symmetry,
$\cH^{\bar e_2,\bar m_2}=\Ext(\cB_\cc^\vee,{}^L\!\cB^{\bar m_2,\bar e_2})$
in the $B$-model on $\MH(C,\LG)$ with $J$ \cite{Wu15,Wu18}.

Like the Hitchin moduli space $\MH(C,G)$, the moduli space $\MH(C',G)$ for
non-orientable $C'$ should also contain $T^*\Mf(C',G)$ as a dense open set
\cite{Wu18}.
Here $\Mf(C',G)$ is the moduli space of flat $G$-connections on $C'$.
The component $\Mf^{m_2}(C',G)$ with topological type $m_2$ is connected
\cite{HL03} and has an action of $\ker(\del_{\bZ_2})$ \cite{Wu15}.
Using the real polarisation of $T^*\Mf(C',G)$, $\cH^{\bar e_2,\bar m_2}$ is
the space of wave functions on $\Mf^{m_2}(C',G)$ that transform according to
$\bar e_2$ under $\ker(\del_{\bZ_2})$.

\bigskip\medskip\noindent{\bf Acknowledgments.}
The author is supported in part by grant No.~106-2115-M-007-005-MY2 from MOST
and the NCTS (physics) of Taiwan.

\end{document}